\documentclass[final]{ustcstep}

\newcommand{\modi}{\textcolor{black}}

\usepackage{lineno}
\usepackage{graphicx}
\usepackage{lscape}
\usepackage{multirow}
\usepackage{color}
\usepackage{url}

\newcommand\addr[2]{{\footnotesize \it $^{#1}$#2}\\}
\usepackage[pdfborder={0 0 0},urlcolor=blue,breaklinks]{hyperref}

\usepackage{enumerate}
\usepackage{natbib}
\usepackage{epstopdf}
\usepackage{caption}
\usepackage{amsmath}
\usepackage{tabularx,booktabs}
\usepackage[figuresright]{rotating}
\usepackage{float}

\begin{document}

\title{Why the Shock-ICME Complex Structure is Important: Learning From the Early 2007 September CMEs}


\author{Chenglong Shen$^{1,2,3,*}$, Mengjiao Xu$^{1}$, Yuming Wang$^{1,4}$, Yutian Chi$^{1}$, and Bingxian Luo$^{5,6}$\\
  \addr{}{$^1$ CAS Key Laboratory of Geospace Environment, Department of Geophysics and Planetary Sciences, }\\
  	\addr{}{University of Science \& Technology of China, Hefei, Anhui 230026, China (clshen@ustc.edu.cn)}\\
	 \addr{}{$^2$ Collaborative Innovation Center of Astronautical Science and Technology, Hefei, 230026, China}\\
	 \addr{}{$^3$ Synergetic Innovation Center of Quantum Information \& Quantum Physics,}\\
	 \addr{}{ University of Science and Technology of China, Hefei, Anhui 230026, China}\\
	 \addr{}{$^4$ Mengcheng National Geophysical Observatory, School of Earth \& Space Sciences,}\\
	 \addr{}{ University of Science \& Technology of China, Hefei, China}\\
     \addr{}{$^5$ National Space Science Center, Chinese Academy of Sciences, No. 1 Nanertiao Zhongguancun Haidian District,}\\
     \addr{}{ Beijing, 100190, China}\\
     \addr{}{$^6$ School of Astronomy and Space Science University of Chinese Academy of Sciences,}\\
     \addr{}{ Yuquan Road, Shijingshan Dirtrict, Beijing, 100049, China}\\
	 \addr{}{$^*$ Corresponding author}}

\maketitle
\tableofcontents

\begin{abstract}
In the early days of 2017 September, \modi{an exceptionally energetic} solar active region AR12673 aroused great interest in the solar physics community.
It produced four X class flares, more than 20 CMEs and an intense geomagnetic storm\modi{, for which the peak value of the Dst index} reached up to -142 nT at 2017 September 8 02:00 UT.
In this work, we check the interplanetary and solar source of this intense geomagnetic storm.
We find that this geomagnetic storm was mainly caused by a shock-ICME complex structure, which was formed by a shock driven by the 2017 September 6 CME propagating into a previous ICME
which was the interplanetary counterpart of the 2017 September 4 CME.
To better understand the role of this structure, we conduct the quantitative analysis about the enhancement of  ICME's geoeffectiveness induced by the shock compression.
The analysis shows that the shock compression enhanced the intensity of this geomagnetic storm \modi{by a factor of two}.
Without shock compression, there would be only a moderate geomagnetic storm with a peak Dst value of $\sim$ -79 nT.
In addition, the analysis of the proton flux signature inside the shock-ICME complex structure shows that this structure
also enhanced the solar energetic particles (SEPs) intensity by a factor of $\sim$ 5.
These findings illustrate that the shock-ICME complex structure is a very important factor in solar physics study and space weather forecast.
\end{abstract}

	
\section{Introduction}

Coronal Mass Ejections (CMEs) are the most \modi{energetic} eruptions from the Sun.
When CMEs continuously erupt from the Sun, they may interact with each other during the propagation from the Sun to 1 AU \citep[e.g.][]{Gopalswamy:2001kh}.
Using the large field view observations from the Heliospheric Imager (HI) on board the Solar TErrestrial RElations Observatory (STEREO)\citep{2008SSRv..136....5K},
the kinematic \modi{evolution} of CMEs have been widely studied \citep[e. g.][and references therein]{Shen2012a,Lugaz2012a,Temmer2014a,2012ApJ...749...57T,Colaninno2015,Mishra2017a,Shen2017a,Lugaz2017,Manchester2017a}.
These results show that the space weather effect of CMEs, such as when and \modi{what CME structures} will \modi{impact} the Earth, \modi{can be greatly effected} during the CME's interaction.

The interaction between multiple CMEs can form complex structures as seen from the in situ measurements.
Such complex structures have been called complex ejecta, or multiple interplanetary CMEs (ICMEs) \citep[e.g.][]{2002JGRA..107.1266B}.
If the ICMEs show obvious characters of magnetic clouds (MC), they are also \modi{referred to} multiple MCs \citep[e.g.][]{Wang2003}.
In addition, when the shock driven by the following CME propagates into the previous CME, they may form a special type of complex structure called shock-ICME (or shock-MC) structure
\citep[e.g.][]{Ivanov1982,Lepping1997,Wang2003,Lugaz2015a,Shen2017}.
In such \modi{structures}, the shock will compress the magnetic field inside the ICME, thereby often enhancing the geoeffectiveness of the ICME, according to previous case \modi{studies} \citep[e.g.][]{Lepping1997,Wang2003a,Lugaz2015b},
analytical study \citep{Wang:2003bq}, magnetohydrodynamic (MHD) simulations \citep[e.g.][]{Vandas1997,Lugaz2005a,2006JGRA..11111102X,Shen:2011bc,Shen2012e}  and statistical analyses \citep{Lugaz2015a,Shen2017}.
For the space weather study and forecasting, the most important thing is the enhancement of ICMEs' geoeffectiveness caused by the shock compression.
Based on a simple theoretical model, \cite{Wang:2003bq} studied the possible effect of \modi{the shock} compression on the geoeffectiveness of the shock-ICME structure,
\modi{and} suggested that 1 time enhancement of $vB_s$ would make the Dst index enhance 1.73 times based on an empirical formula relating the Dst index to the interplanetary parameters.

\modi{In addition}, Solar Energy Particle (SEP) \modi{events are} another important space weather \modi{phenomena} that may be affected by the interaction between shock and ICME.
\cite{Shen2008} reported that the proton flux was significantly enhanced in the shock-MC structure in the 2001 November 5 event,
which differed from the normal picture that the proton flux would \modi{decrease} in isolated ICME structures \citep[e.g.][and references therein]{Cane:2006kg}.
This \modi{enhancement} might be due to the combined effects of the shock and the MC boundaries: the shock can accelerate particles within the MC and the MC boundaries prevent the leakage of these accelerated particles.
It is worth noting that the enhancement in the shock-MC structure in the 2001 November 5 event is \modi{associated with} the largest SEP \modi{event} in solar cycle 23.

In the early days of 2017 September, the active region AR12673 passed \modi{across} the visible side of the Sun.
This \modi{extremely energetic} active region \modi{produced} more than 80 flares including four X class flares within 7 days.
Two in four X class flares, X9.3 and X8.2 flares\citep[e.g.][]{Yan2018,Yan2018a} , are ranked as the top two flares in solar cycle 24 till now.
AR12673 also produced more than 20 CMEs from 2017 September 4 to September 10.
Thus, we can expect that these CMEs may interact with others during their propagations from the Sun to the Earth.
In addition, an intense geomagnetic storm occurred on 2017 September 8 02:00 UT with a peak Dst value ($Dst_{min}$) of -142 nT according to the real time Dst observation provided by World Data Center (WDC).
Based on the in situ observations by Wind and Deep Space Climate ObseRvatory (DSCOVR) spacecraft near the Earth, an obvious shock-ICME complex structure \modi{was} the main source of this intense geomagnetic storm.
\modi{Thus, in the paper,} we will mainly focus on the space weather effect of shock-ICME complex \modi{structures in this period}.
The detailed in situ observations of ICMEs and their solar sources will be shown in Section 2.
In Section 3, we will quantitatively discuss the significance of the shock-ICME complex structure in generating the geomagnetic storms.
The influence of the shock-ICME complex structure on SEP intensity will be investigated in Section 4.
We will give the conclusions and make some brief discussions in the last section.


\section{Interplanetary and solar sources of the geomagnetic storms}

\begin{figure}
        \centering
        \noindent\includegraphics[width=0.8\linewidth]{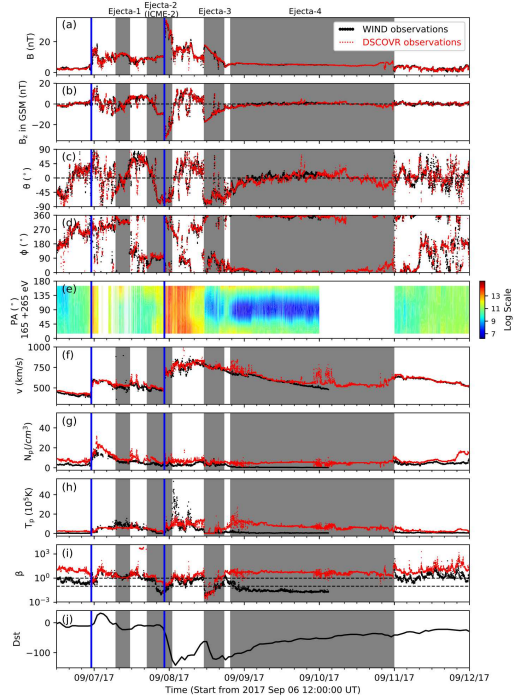}
        \caption{The Wind and DSCOVR in situ observations from 2017 September 6 to September 13. The black symbols show the WIND observations while the red symbols show the DSCOVR observations.
                 From top to the bottom, panels are the magnetic field strength ($B$),  north component of the magnetic field in GSM coordinate system ($B_z$), the elevation ($\theta$) and azimuthal ($\phi$) of magnetic field direction in GSM coordinate system, solar wind speed ($v$), proton density ($N_p$), proton temperature ($T_p$) and the ratio of proton thermal pressure to magnetic pressure ($\beta$) and the $Dst$ from World Data Center (WDC). Shade regions show the period of the ICMEs while the blue vertical lines show the time of shocks. }
        \label{insitu_fig}
\end{figure}

To check the geomagnetic activity and the possible interplanetary \modi{drivers} of the geomagnetic storm, in situ observations from 2017 September 6 to September 11 from the Wind (black symbols) and DSCOVR (red symbols) spacecrafts
as well as the Dst observations from WDC are shown in Figure \ref{insitu_fig}.
Seen from this figure, there was an intense and multi-step geomagnetic storm. The peak of the storm occurred at 2017 September 8 02:00 UT with a value of -142 nT.
The sudden commencement of this storm occurred at 2017 September 7 01:00 UT after the first shock (shown as the first vertical blue line) arrived at the Earth.
Soon later, the Dst index began to decrease when the sheath region and the ejecta region (Ejecta-1) of the first ICME (called as ICME-1 hereafter) hit the Earth.
The leading edge of Ejecta-1 arrived at the Earth at 2017 September 07 06:50 UT and the trailing edge of it arrived at 2017 September 07 11:30 UT (shown by the first shaded region) .
During this period, the in situ observations exhibited obvious signatures of \modi{a} magnetic cloud (MC) with enhanced magnetic field intensity, smooth rotated magnetic field vector, low temperature and low plasma beta.
After the passage of ICME-1, the Dst index began to recover.
About 10 hours later, the Dst index started to decrease again when the second ejecta (shown by the second shaded region as Ejecta-2/ICME-2) hit the Earth.
The ICME-2 started at 2017 September 07 16:50 and ended at 2017 September 08 01:00 UT.
Meanwhile, Wind and DSCOVR recorded a shock at 2017 September 7 22:28 UT, before the trailing edge of ICME-2.
It means that this shock  was propagating inside the ejecta region of ICME-2 and formed a shock-ICME complex structure.
Seen from panel (j) in Figure \ref{insitu_fig}, the Dst index decreased quickly since the arrival of the shock inside ICME-2.
This sudden and quick decrease owed to the large south component of the magnetic field in the ejecta region of ICME-2 which was compressed by the shock.
At 2017 September 8 02:00 UT, the Dst index reached its peak value of -142 nT.
Thus, this intense geomagnetic storm was mainly caused by the shock-ICME complex structure.
After that, two other ejecta were observed near the Earth.
They are marked as the 3rd and 4th shaded regions in Figure \ref{insitu_fig}.
The third shaded region (called as Ejecta-3 hereafter) has the signature of enhanced magnetic field, low proton temperature, and bi-direction electron beam.
But, no obvious rotation of the magnetic field vector can be found. All the magnetic field carried by this structure \modi{points} southward.  This structure caused another Dst peak with the value of -124 nT.
The 4th shaded region in Figure \ref{insitu_fig} is a long lasting ejecta (\modi{referred to} Ejecta-4 hereafter) from 2017 September 08 19:30 UT to September 11 00:00 UT.

\begin{figure}
\centering
\noindent\includegraphics[width=0.7\linewidth]{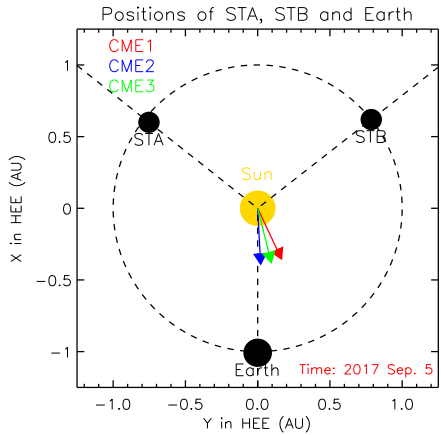}
\caption{Positions of STEREO A, Earth, STEREO B at the time of 2017 September 5. Different color arrows show the propagation directions of different CMEs.}
\label{pos}
\end{figure}

\begin{figure}
\centering
\noindent\includegraphics[width=0.8\linewidth]{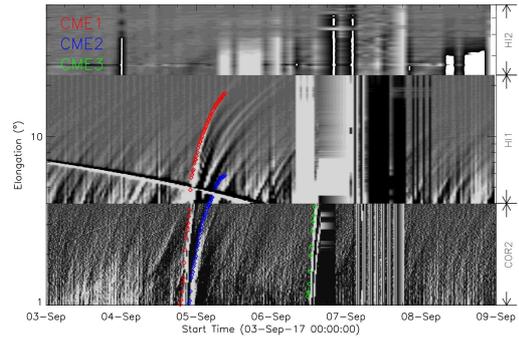}
\caption{Time-elongation angle map from 2007 September 4 to September 9 based on STEREO A observations. \modi{Different colors show the measurements of the front edges of different CMEs.}}
\label{jmap}
\end{figure}

Overall, there are 2 shocks and 4 ejectas recorded by the in situ measurements near the Earth from September 6 12:00 UT to September 11.
The shock times, begin and end times of these ejectas are shown in rows 2 - 4 of Table \ref{icme}.
In order to find the possible solar sources of these structures, \modi{we further check the coronagraph observations of STEREO-A and SOHO 2017 September 3 to September 8.}
Figure \ref{pos} show the \modi{relative} position of Earth and STEREO satellites.
During this period, the separation angle between Earth and STEREO A is $\sim$ 128 $^\circ$. Thus, an Earth-directed CMEs can be well observed by the STEREO-A.
Figure \ref{jmap} shows the Time-Elongation Angle map from 2017 September 3 to 2017 September 8.
A 64-pixel-wide slice is placed along the ecliptic plane in the running-difference images from COR2, HI1 and HI2 onboard STEREO-A to produce this J-map.
Seen from this figure, three different trajectories, which \modi{correspond} with three CMEs, \modi{can} be well observed.
Meanwhile, the coronagraph images from STEREO-A and SOHO also show that there \modi{are} three Earth-directed CMEs erupted from the Sun during this period.

\begin{figure}
\centering
\noindent\includegraphics[width=0.8\linewidth]{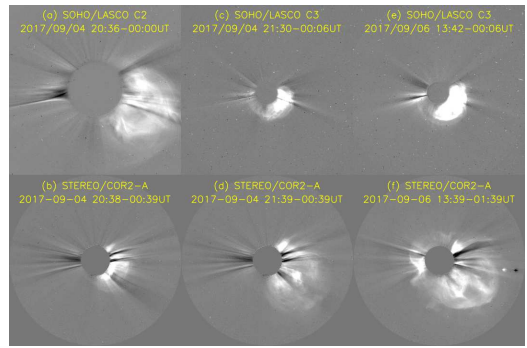}
\caption{SOHO/LASCO and STEREO/SECCHI observations the 2017 September 4 19:48 UT (panel (a) and (c)), 2017 September 4 20:36 UT (panel (b) and (e)) and 2017 September 6 12:24 UT (panel (c) and (f)) CMEs.}
\label{cme}
\end{figure}

Figure \ref{cme} shows the coronagraph images of these CMEs. The first CME (CME-1) was \modi{first} observed by STEREO A \modi{at} 2017 September 4 18:54 UT \modi{and was first observed by SOHO/LASCO at} 2017 September 4 19:00 UT.
The front edge of this CME \modi{creates} the first track in the J-map shown as the red symbols in Figure \ref{jmap}.
About 1 \modi{hour} later, another Earth-directed CME (CME-2) appeared in the SOHO and STEREO-A field of view.
Panels (c) and (d) show the coronagraph observations of this CME. This CME was first observed by STEREO A at 2017 September 4 19:54 UT and was first observed by SOHO at 2017 September 4 20:36 UT.
Seen from SOHO, this is a halo CME. The second trajectory in Figure \ref{jmap} , which \modi{is} indicated by blue symbols, \modi{shows} the position of the front edge of this CME.
Another Earth-directed CME (CME-3) was observed by SOHO and STEREO-A about two days later.
At 2017 September 6 11:54 UT, this CME was first observed by STEREO-A/COR2 at its south west direction.
Half \modi{an hour} later, this CME was observed by SOHO/LASCO at 2017 September 6 12:24 UT.
This CME was associated with a X9.2 class flare \modi{and appeared} a halo CME in the SOHO/LASCO observations.
Green symbols in Figure \ref{jmap} show the position of its front edge and panels (e) and (f) show the coronagraph images of this CME.
The 5th column in Table \ref{icme} shows the time when these CMEs were first observed by SOHO/LASCO.

\begin{figure}
\centering
\noindent\includegraphics[width=0.8\linewidth]{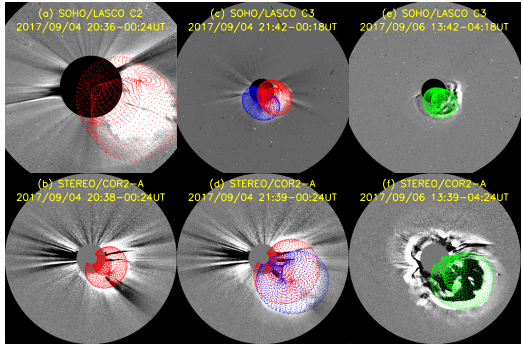}
\caption{The coronagraph images of these CMEs with GCS wireframe overlaid on top. Red, blue and green colors show the fitting results for CME-1, CME-2 and CME-3, respectively.}
\label{gcs}
\end{figure}

In addition, the graduated cylindrical shell (GCS) model, which was developed by \cite{Thernisien2006,2009SoPh..256..111T,Thernisien:2011jy}, \modi{is} applied to get the 3 dimensional parameters of these CMEs.
Figure \ref{gcs} shows the fitting results of these CMEs. Seen \modi{from} these images, the GCS model can well represent the topology of these CMEs.
The last three columns in Table \ref{icme} show the fitting results of these CMEs, including the propagation directions, velocities and face-on angular widths.
Assuming a constant velocity and considering the influence of the propagation direction and angular width on the prediction of the arrival time suggested by \cite{Shen2014a},
CME-1 would arrive at the Earth around the time of September 6 22:27 UT. However, previous results show that fast CME would decelerate during \modi{their} propagation in interplanetary space\citep[e.g.][and reference therein]{2001JGR...10629207G,Vrsnak2001,Vrsnak2007d,Gopalswamy2005a,Temmer2011,Lugaz:2012es}.
Such deceleration may make the CME-1 arrive at the Earth later than September 6 22:27 UT.
Thus, CME-1 \modi{is} more likely to be the solar source of ICME-1.
Seen from the Figure \ref{jmap}, the front edge of the CME-2 \modi{is} lower than the front edge of CME-1\modi{ indicating} that CME-2 would arrive at the Earth later than CME-1.
Thus, CME-2 was the solar source of the ICME-2.
It should be noted that, based on the \modi{fitting results of GCS model}, CME-2 \modi{is} faster than CME-1 and their propagation directions \modi{are} close to each other.
Thus, these two CMEs \modi{are} expected to be interacted in the interplanetary space.
Seen from the in situ observations, possible interaction region signatures \modi{were} detected between these two ICMEs with lower magnetic field, higher velocity, high density and higher plasma beta.
Furthermore, similar analysis shows that CME-3 might arrive at Earth after September 7 19:21 UT. Considering the long duration of Ejecta-4 and the larger angular width of CME-3,
\modi{we verify that}  CME-3 \modi{is} the solar source of ICME-4 and \modi{the driver} of the second shock.
It should be noted that, no obvious Earth-directed CME could be identified as the solar source of Ejecta-3.
A possible explanation is that this ejecta structure is formed in the sheath region of ICME-4 during \modi{its} propagation outward\citep[e.g.][and reference therein]{Zheng2018}.


\begin{table*}
\scriptsize
\begin{center}
        \caption{The list of the ICMEs or ICME-like structures and their solar sources from 2017 September 6  to 2017 September 9.} \label{icme}
\begin{tabular}{|c|c|c|c|c|c|c|c|}
        \hline
        No   &Shock Arrival (UT)  & Begin  (UT)  & End  (UT) & CME Time (UT)$^1$ &Propagation Direction & Velocity (km/s) & Face-on Width ($^\circ$) \\
        \hline
        1&Sep. 6 23:06  & Sep. 7 06:50 & Sep. 7 11:30 & Sep. 4 19:00 & S08W25 & 1005 & 73 \\
        \hline
        2& --- & Sep. 7 16:50   & Sep. 8 01:00  &  Sep. 4 20:24 & S25W03  & 1766 & 75 \\
        \hline
        3& ---  & Sep. 8 11:05   & Sep. 8 17:38  &    & & & \\
        \hline
        4&Sep. 7 22:28  & Sep. 8 19:30   & Sep. 11 00:00 & Sep. 6 12:24 & S18W14 & 1548 & 80 \\
        \hline
\end{tabular}
$^1$ The time of this CME was first observed by SOHO/LASCO.
\end{center}
\end{table*}

\section{The Importance of Shock-ICME Complex Structure in Causing the Geomagnetic Storm}

\begin{figure}
\centering
\noindent\includegraphics[width=0.8\linewidth]{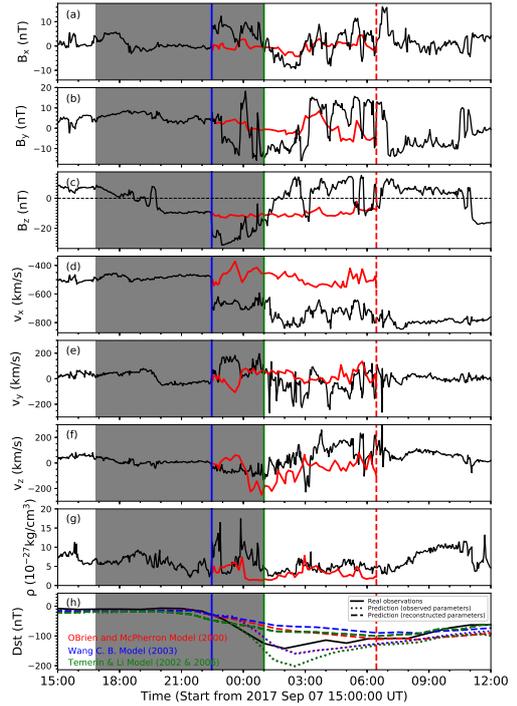}
\caption{\modi{The observational data and recovered uncompressed state of magnetic field, solar wind speed, total plasma density and Dst index from 2017 September 6 12:00 to 2017 September 8.
The shade region shows the period of the ICME and the blue line shows the time of the shock arrival.
The black lines in panel (a) to (h) between the first two vertical lines (blue and green vertical lines) show the original observations, and the red lines between the first and third vertical lines (blue and red vertical lines)
represent the recovered parameters.
Panel (h) shows the real data (black line) and the prediction results based on the observed (dashed lines) and recovered (dashed-dotted lines) parameters of Dst index. Different colors represent different prediction methods.}}
\label{icme_rec}
\end{figure}

\begin{figure}
\centering
\noindent\includegraphics[width=0.8\linewidth]{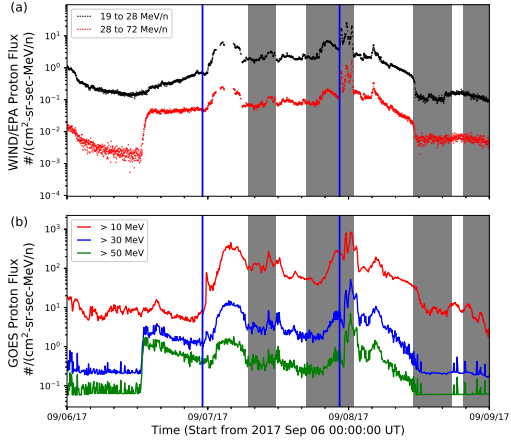}
\caption{The observations of the flux of high energy protons from Wind/EPA \modi{(panel (a))} and GOES \modi{(panel (b))} during the period from 2017 September 6 to 2017 September 8. \modi{Different colors represent different energy channels. Four shaded regions show four ICMEs and two blue vertical lines show the shock positions.}}
\label{proton}
\end{figure}

Based on the observational analysis in Section 2, we \modi{find} that the intense geomagnetic storm with the peak value of -142 nT is caused by the \modi{shock-ICME structure}.
Seen from Panel (b) in Figure \ref{insitu_fig}, the $B_s$, which is equal to the negative value of $B_z$, jumps from $\sim$ 10 nT to $\sim$ 30 nT at the shock inside ICME2.
As the $B_s$ or the dawn-dusk electric field ($v_xB_s$) is a main factor in determining the intensity of a geomagnetic storm\cite[e.g.][]{Gonzalez1994},
we can expect that such enhancement of $B_s$ do significantly enhance the intensity of this geomagnetic storm.
But, the question is: how \modi{much} the shock enhances the geoeffectiveness of this ICME?

Recently, \cite{Wang2018} developed a method to recover the shocked part of the magnetic cloud to the uncompressed state.
One can refer to Section 3.2 of their paper for details. \modi{For} the completeness of the paper, we repeatedly describe the method below.
This method simply assumes that: (1) the magnetic field, plasma velocity and density in the sheath region can be related to the uncompressed state by the shock relation, i.e., Rankine-Hugoniot jump conditions, and (2) the shock normal (\^{n}), shock speed ($v_s$), and the compression ratio ($r_c$), are the same as those at the observed shock surface.
Treating the sheath region as the downstream (using subscript `2') of the shock,
the uncompressed state, i.e., the parameters of the upstream (using subscript `1') of the shock, can be given by the following equations:
\begin{eqnarray}
\rho_1=\frac{1}{r_c}\rho_2 \\
\mathbf{B_{1n}}=\mathbf{B_{2n}}\\
\mathbf{B_{1\perp}}=\frac{v_{A2}^2-u_2^2}{v_{A2}^2-r_2u_2^2}\mathbf{B_{2\perp}}\\
\mathbf{u_{1\perp}}=r_c\mathbf{u_{2\perp}}\\
\mathbf{u_{1\perp}}=\frac{v_{A2}^2-u_2^2}{v_{A2}^2-r_2u_2^2}r_c\mathbf{u_{2\perp}}
\end{eqnarray}

in which $\rho$ is the density including the protons and electrons,
$B$ represents the magnetic field,
$u$ is the solar wind speed in the DeHoffman-Teller (HT) frame, and $v_A$ is the Alfv\'{e}n speed.
The subscript `$n$' and '$\perp$' \modi{mean} the component parallel and perpendicular to the shock normal.
The recovered interval is longer than the observed shocked interval, and its duration is calculated  based on the mass conservation with the formula of $dt_1 = \frac{u_{2n}+vs}{u_{1n}+vs}r_cdt_2$.
The shock parameters can be obtained based on the Rankine-Hugoniot (R-H) analysis \citep[e.g.][and references therein]{Koval2008}.
In this case, the parameters of shock inside ICME2 are: shock normal (\^{n}) direction in GSM coordinate = [-0.83,0.3,-0.46]
, shock speed $v_s$ = 759 km/s and compression ratio $r_c$ = 2.23.

Figure \ref{icme_rec} shows the Wind observations of the magnetic field and solar wind velocity vectors (black lines) and the recovered uncompressed state of these parameters (red lines) of the Shock-ICME complex structure (ICME-2).
The shaded region shows the observed shock-ICME region.
The black lines between the first two vertical lines show the original \modi{observations}, and the red lines between the first and third vertical lines \modi{represent} the recovered parameters.
Seen from this figure, the intensity of the magnetic field and the solar wind velocity would became much smaller if there was no shock compression.
Using the recovered state of the ICME, we can then quantitatively estimate the \modi{enhancement of ICME's geoeffectiveness}.
Previous studies show that the peak value of the Dst index \modi{is} well correlated with the value of the $B_s$ and the dawn-dusk electric field ($v_xB_s$)\citep[e.g.][and references therein]{Wang2003b,Wu2016,Shen2017}.
Based on the in situ observation, the peak \modi{values} of $B_s$ and $v_xB_s$ in this ICME were 31 nT and -13 mv/m respectively.
But, if the shock compression did not happen, the peak \modi{values} of $B_s$ and $v_xB_s$ would decrease to 13 nT and -6.4 mv/m, according to the recovered uncompressed results.
\cite{Shen2017} have shown a statistical correlation between the $v_xB_{s,min}$ and $Dst_{min}$. It is $Dst_{min}=8.48(v_xB_s)_{min} (mv/m)-24.5$.
Based on this correlation, we can estimate that the value of $Dst_{min}$ is -135 nT by using observed $v_xB_s$ or -79 nT by using the recovered $v_xB_s$.
It can be seen that, the calculated value of the $Dst_{min}$ from the observational solar wind data is similar with the real observed value (-142 nT).
Besides, without the shock compression, the peak value of the $Dst$ index would decrease greatly. The possible peak value is -79 nT, which is larger than -100 nT.
Thus, there would only be a moderate geomagnetic storm if the ICME was not compressed by the shock.
Using the peak Dst index as a measure, we can calculate that the shock compression enhanced the intensity of the geomagnetic storm \modi{by a factor of 2} ($\sim$ 1.8).
In addition, \modi{other} Dst index forecasting models are applied \modi{to} the observed and recovered solar wind parameters. Panel (h) in \modi{Figure} \ref{icme_rec} shows the result.
The black line shows the real \modi{observation} of the Dst index. The dashed lines and dashed-dotted lines show the \modi{magnetosphere} prediction results based on the observed and recovered parameters, respectively.
Red, blue and green lines show the results from OBrien and McPherron (OM) model \citep{OBrien2000}, Wang model \citep{Wang2003c} and Temerin and Li (TL) model \citep{Temerin2002}, respectively.
Seen from this panel, all these models can well predict the tendency of the Dst variation by using the real solar wind parameters.
But, the predicted peak values of Dst are all higher than real observations. They are -158 nT (OM model), -160 nT (Wang model) and -202 nT (TL model).
Meanwhile, by using the recovered parameters, the predicted Dst index decrease much slower and the predicted peak values of Dst index are -112 nT, -91 nT and -101 nT for \modi{the respective} models.
Compared with prediction results using the real solar wind observations, ratios between the peak values of Dst are of 1.4, 1.8 and 2 for different models, respectively.
Thus, combined with these results, we suggest that the shock compression enhanced the geoeffectiveness of ICME-2 \modi{by an average of 1.73}.
Without shock compression, ICME-2 would only cause a moderate geomagnetic storm.

\section{The Proton Flux Signature in the Shock-ICME structure}
The proton flux enhancement during a shock-MC structure in 2011 November 5 event has been reported by \cite{Shen2008}.
In the present study, the characteristics of high energetic proton flux are also \modi{provided}.
Figure \ref{proton} shows the high energy proton observation from Wind/EPA (panel (a)) and GOES (panel (b)). Different colors represent different energy channels.
Seen from these panels, the proton flux decreased at the front edges of the shaded regions 1 and 3.
Meanwhile, it increased at the trailing \modi{edges} of these regions.
This indicates that the proton flux intensities in the Ejecta-1 and the Ejecta-3 are lower than those in background.
This is consistent with the normal situation\citep[e.g.][and references therein]{Cane:2006kg}.
However, for the shaded region 2 which is the shock-ICME complex structure, the proton flux intensities increased at its front edge and decreased at its trailing edge for almost all energy channels.
At the front edge, the intensity \modi{of} energy $\ge 10$ MeV protons jumped about 5 times from $\sim$40 pfu to $\sim$200 pfu. In addition, the arrival of the shock \modi{makes} this intensity further enhanced.
At the trailing edge of this region, the proton flux intensity decreased about 10 times from $\sim$800 pfu to $\sim$80 pfu.
It means that the proton flux intensity in this structure was obviously higher than that in the background.
In addition to the 2001 November event reported by \cite{Shen2008}, this is another definitive case that the shock-ICME complex structure leads to a significantly enhancement in the proton flux intensity.

\section{Conclusion and Discussion}
In this work, we studied the interplanetary signature and the cause of the intense geomagnetic storm in the early days of 2017 September in detail.
Based on the in situ observations, we found that there were three obvious ICMEs and one ICME-like structure.
Two ICMEs drove shocks ahead of them.
It is noteworthy that the shock driven by following structure propagated into ICME-2 and then formed a shock-ICME complex structure.
The space weather effect of these ICMEs especially the shock-ICME complex structure \modi{was} further discussed. The main findings of this work are:
\begin{enumerate}
\item These ICMEs caused a multiple step intense geomagnetic storm with the peak value of the Dst index of -142 nT.
The shock-ICME complex structure formed by ICME-2 and shock driven by ICME-3 was the main interplanetary cause of this intense geomagnetic storm.

\item Using the recovering method developed by \cite{Wang2018}, we showed that the shock compression in ICME-2 obviously enhanced the magnetic field and also the geoeffectiveness of this ICME.
A quantitative analysis showed that this ICME would only cause a moderate geomagnetic storm if no compression happened.
The compression of the magnetic field by the shock made the intensity of this geomagnetic storm enhanced \modi{by roughly a factor of two}.

\item The high energy proton flux in this shock-ICME complex structure was obviously enhanced, which was similar to another shock-ICME event reported by \cite{Shen2008}.
The proton flux intensity in the shock-ICME complex structure was about 5 times higher than that in the background, which make this SEP event stronger.

\end{enumerate}

In this work, we showed the \modi{enhancement} of the ICMEs' geoeffectiveness caused by the shock compression quantitatively based on the observations for the first time.
Our results showed that the ICME-2 would only \modi{cause} a moderate geomagnetic storm without the compression of the ICME by the shock.
Meanwhile, we also found that the proton flux intensity enhanced in this shock-ICME complex structure.
These findings further strength the viewpoint that the multiple CMEs interaction especially the shock-ICME interaction is an important factor in the space weather effect of CMEs.
But, questions still remain.  The first question is how we forecast the shock-ICME interaction based on the solar observations of CMEs?
The second question is what parameters would influence the geoeffectiveness and SEP intensity enhancement?
Especially, we reported two proton flux enhancements events due to the shock-ICME complex structures. Does this occur in all shock-ICME complex structures?
To answer these questions, more detailed and statistical analysis should be done.

\acknowledgments
The authors thank the referee for the comments that helped to improve this paper.
We acknowledge the use of the data from \textit{SOHO}, \textit{STEREO}, \textit{Wind}, \textit{DSCOVR} and \textit{GOES} satellites and the world data center  (WDC) for Geomagnetism, Kyoto.
This work is supported by grants from CAS (Youth Innovation Promotion Association CAS and Key Research Program of Frontier Sciences QYZDB-SSW-DQC015),
NSFC (41774181,41774178, 41574165, 41474164, 41761134088), the Fundamental Research Funds for the Central Universities (WK2080000077) and the Specialized Research Fund for State Key Laboratories.


\end{document}